\documentclass[floatfix,prb,twocolumn,showpacs]{revtex4}
\usepackage{dcolumn}
\usepackage{graphicx}
\usepackage{epstopdf}
\usepackage{color}
\usepackage{amssymb}
\usepackage{amsmath,epsf}
\begin{document}
\author {Saad Elgazzar$^{1,2}$, A. M. Strydom$^1$, and Stefan-Ludwig Drechsler$^3$}
\affiliation{
$^1$Physics Department, Univ. of Johannesburg, P.O. Box 524, Auckland Park 2006, South Africa\\
$^2$Department of Physics, Faculty of Science, Menoufia University, Shebin El-kom, Egypt\\
$^3$Leibniz Institute for Solid State and Materials Research IFW–Dresden, P.O. Box 270116, D-01171 Dresden, Germany}
\title {Full Relativistic Electronic Structure and Fermi Surface Sheets of the First Honeycomb-Lattice Pnictide Superconductor SrPtAs}
\date{\today}
\begin{abstract}
We report full-potential density functional theory (DFT)-based {\it ab initio} band structure calculations to investigate electronic structure properties of the first pnictide superconductor with a honeycomb-lattice structure: SrPtAs. As a result, electronic bands, density of states, Fermi velocities and the topology of the Fermi surface for SrPtAs are obtained. These quantities are discussed in comparison to the first available experimental data. Predictions for future measurements are provided.
\end{abstract}
\maketitle
\section{Introduction}
\label{intro} 
\begin{figure}[bt!] 
\includegraphics[width=.30\textwidth]{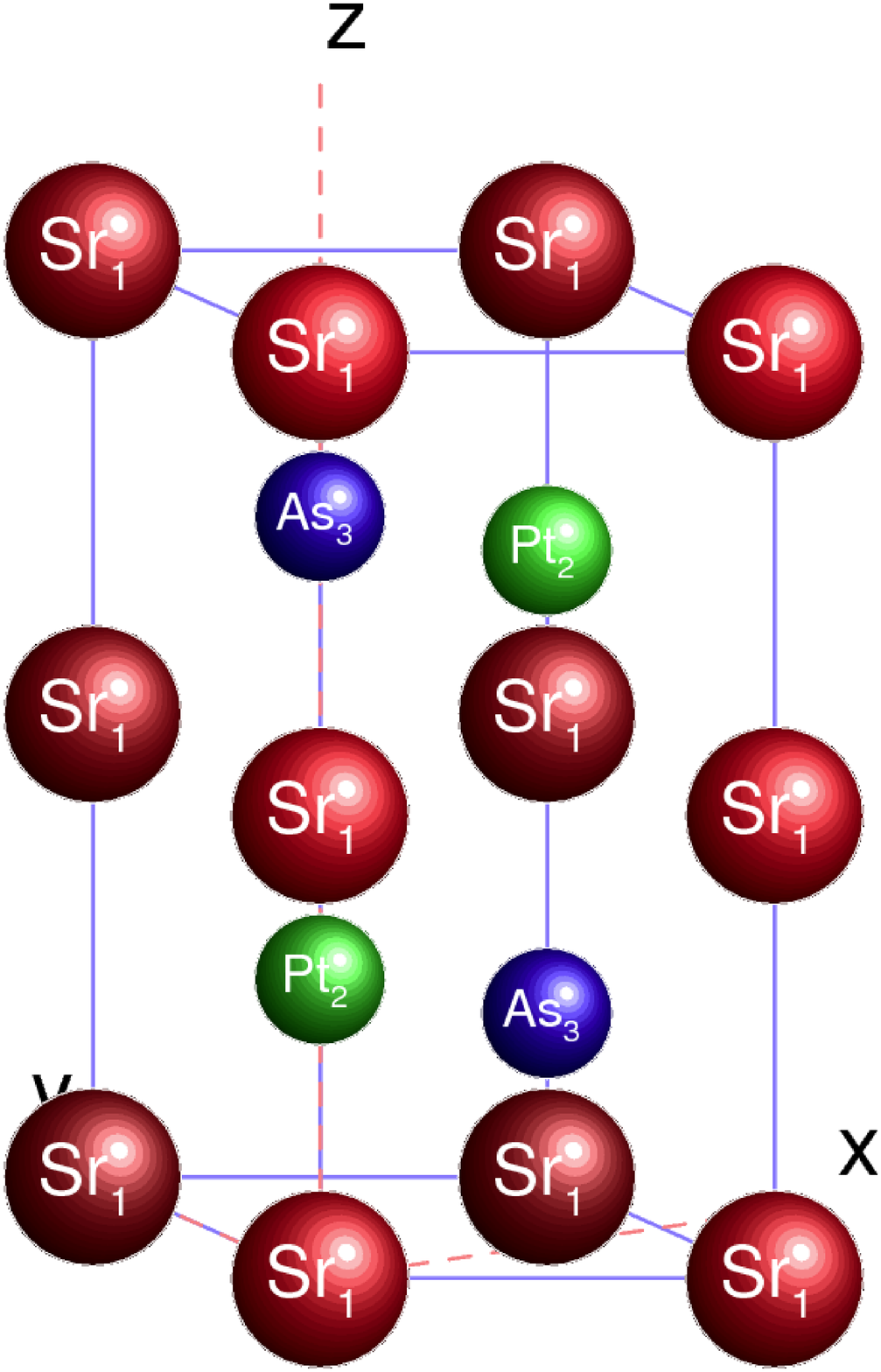} 
\includegraphics[width=.45\textwidth]{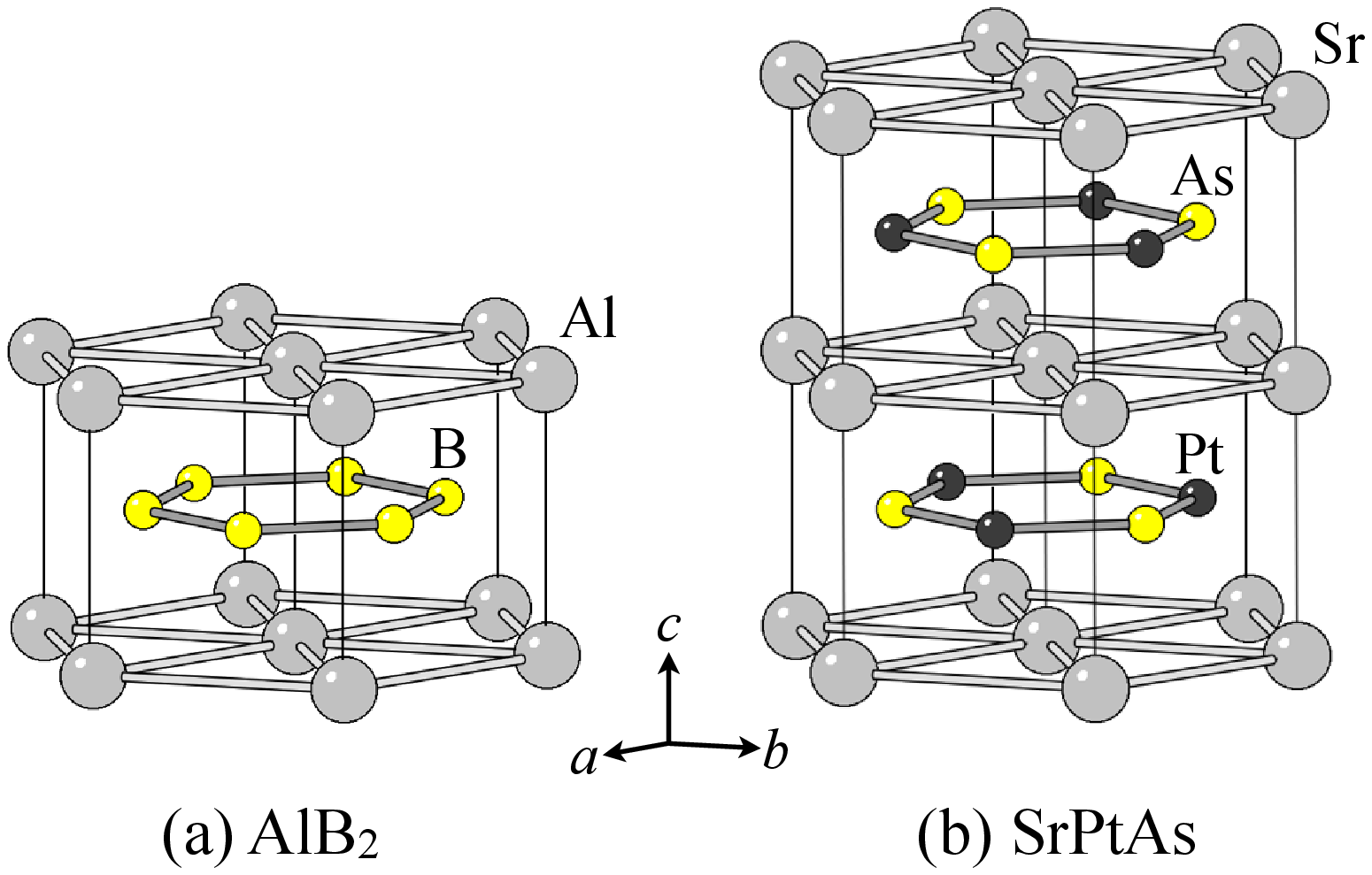} 
\caption{(Color online) The crystal structure of SrPtAs with space group P63/mmc (left and right). The middle and the right panel are taken from Ref.\ 1.} \label{fig1} 
\end{figure}

PtAs layered structures with various local coordinations found considerable interest in the search for novel iron pnictide related superconductors.\cite{Kudo11,Shein11} Special interest is caused by the Fe free SrPtAs and SrPt$_2$As$_2$ systems yielding $T_c$-values of of $2.4~K$ and $5.2~K$, respectively. In the latter superconductivity occurs near a CDW-state pointing to a non-neglible electron--phonon interaction. Ca$_{10}$(Pt$_4$As$_8$)(Fe$_{2-x}$Pt$_x$As$_2$)$_5$ \cite{Shein11a,Kakiya11,Ni11,Cho11,Neupane11,Shulga04} is noteworthy since there in spite of the sizable Pt doping and metallic PtAs-block layers in between the usual FeAs-layers, a large $T_c\approx$ 38~K has been achieved. Here we restrict ourselves to considering the simplest system among that novel class: namely SrPtAs with a honeycomb structure reminiscent of the structure found in  MgB$_2$ \cite{Nagamatsu} (see Fig. \ref{fig1}. There is a clear element of two dimensionality to the structure, with layers of Sr atoms separated by Pt and As atoms arranged in closed-spaced pairs. As compared with an earlier calar relativistic (SR) calculation \cite{Shein11}, we consider here the more precise full relativistic full-potential (FRFP) calculation based on density functional theory (DFT). 
In order to quantify empirically "later on" the role of many-body effects, we also present Fermi Surface (FS) calculations. 
\begin{figure*}[bt!]
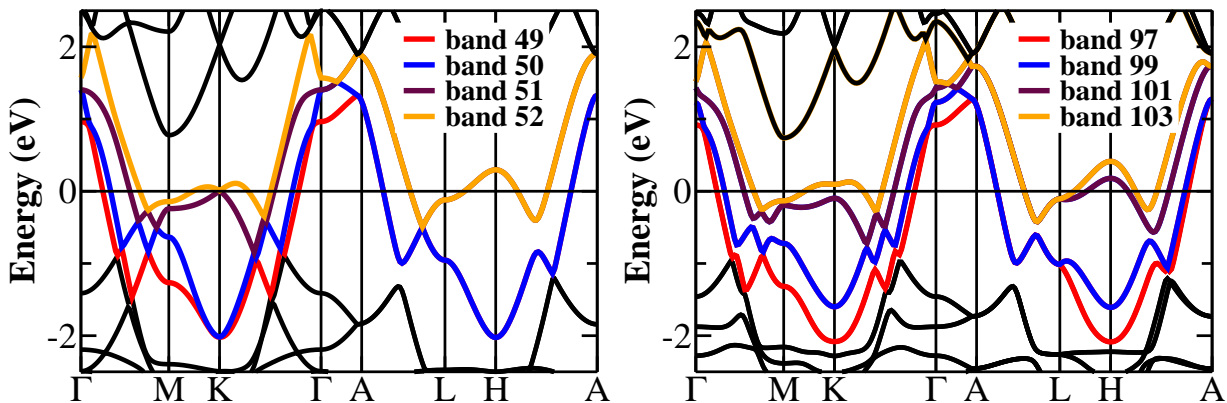

\includegraphics[width=.45\textwidth]{Fig2a.eps}
\includegraphics[width=.45\textwidth]{Fig2b.eps}
\caption{(Color online) The calculated energy bands of SrPtAs within
 the LDA scalar relativistic (SR)[left] and the full relativistic full-potential (FRFP) [right] schemes. The bands crossing the Fermi energy are highlighted highlighted by color.}
\label{fig2}
\end{figure*}

\section{Computational approach}
\label{sec:1}
 Our calculations were carried out employing the fully relativistic version \cite{Opahle,Eschrig,Koepernik} of the full-potential (FRFP)  local orbital (FPLO) minimum-basis band-structure method. In this scheme the 4-component Kohn--Sham--Dirac equation, which  contains the spin-orbit coupling up to all orders,  is solved self-consistently.
 Thereby we adopted the following basis set for the valence states: the $ 4s4p; 5s5p4d$ states for Sr, while for Pt and As we used $ 5s5p; 5d6s6p,$ and $ 4s4p4d$. The high-lying $s$ and $p$ semicore states of Sr and Pt, which might hybridize with the valence states, are in this way included in our basis. The site-centered potentials and densities were expanded in spherical harmonic contributions up to $l_{max} = 12$. The number of $k$-points in the irreducible part of the Brillouin zone was 196, but calculations were performed also with 405 and up to 2176 $k$-points to resolve the fine structure of the density of states at the Fermi energy ($E_F$). We used the Perdew-Wang \cite{Perdew1} parameterization of the exchange-correlation potential in the local spin-density approximation (LSDA).

\section{Structural optimization}
\label{sec:2}
SrPtAs crystallizes in a hexagonal structure derived from the binary AlB$_2$-type as shown in Fig. \ref{fig1}. In SrPtAs, the Al sites are occupied by Sr ions  and the B sites by either Pt or As atoms so that they alternate in the honeycomb layer as well as along the $c$-axis. We have performed {\it ab initio} optimizations of the equilibrium volume and the lattice parameters, $a$ and $c$, using the LDA-relativistic approach. We have computed total energy versus unit-cell volume. The LDA minimum volume slightly underestimates the  experimental value by ~2.4 \%. Similarly, the calculated/optimized lattice constants $a$ and $c$ are in reasonable agreement with the experimental values, i.e.\ $a=4.212$ \AA \  (-0.75\%) and $c=8.753$ \AA \ (-2.63\%). The  small negative deviations of few percents are typical for present day DFT-LDA calculations (see below). Noteworthy is that our optimized lattice constants are closer to the experimental data than those reported in Ref.\ 2 using SR: $a=4.2976$ \AA \ (+1.26\%) and $c=9.0884$ \AA (+1.1\%) which, in contrast, opposite slightly overiiestimate them. 

\section{Band structure}
\label{sec:3}
We have performed nonmagnetic band-structure calculations for the experimental lattice parameters, which are: a=4.244 Å and c=8.989 Å and with internal parameters as Sr (0 0 0), Pt (1/3 2/3 1/4), and As (2/3 1/3 1/4) \cite{Yoshihiro}. The band structure of SrPtAs computed within the SR FRFP schemes are presented in Fig. \ref{fig2}. We find for the SrPtAs system using either the SR or FRFP scheme four bands which cross the Fermi surface.

\begin{figure*}[bt!]
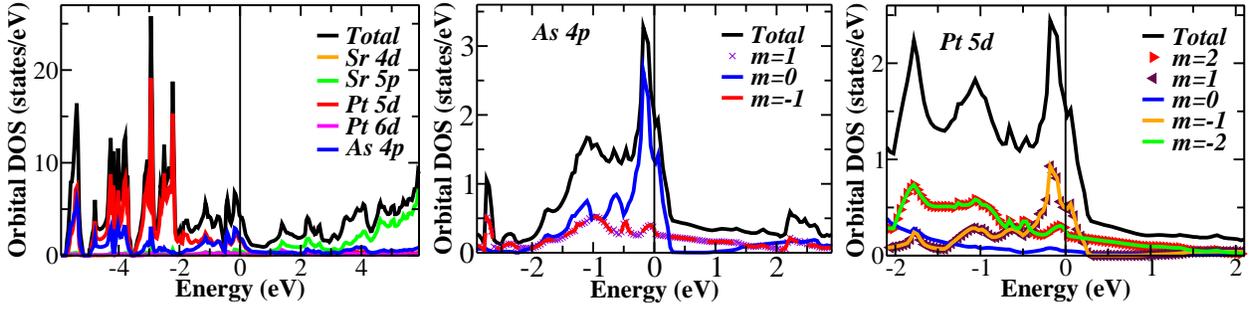

\includegraphics[width=.31\textwidth]{Fig3a.eps}
\includegraphics[width=.3\textwidth]{Fig3b.eps}
\includegraphics[width=.3\textwidth]{Fig3c.eps}
\caption{(Color online) Orbital and partial DOS for SrPtAs, calculated within the LDA-full relativistic full-potential (FRFP) approach.}
\label{fig3}
\end{figure*}
\begin{figure*}[bt!]
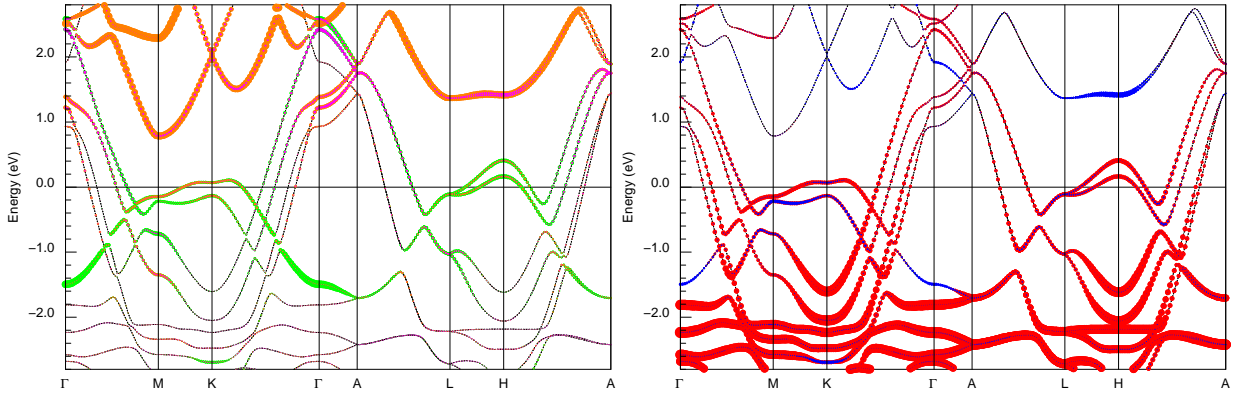

\includegraphics[angle=-90,width=.45\linewidth]{Fig4a.epsi}
\includegraphics[angle=-90,width=.45\linewidth]{Fig4b.epsi}
\caption{(Color online) The  bands of SPtAs  in the vicinity of the Fermi level.
The Pt 5$d$, As 4$p$ and Sr 4$d$ characters are indicated by red, green, and blue symbols, respectively, and their relative weights are given by thickness of the curves. Left: scalar relativistic (SR) calculations. Right: Full relativistic full potential (FRFP) calculations.}
\label{fig4}
\end{figure*}

These bands are denoted as bands 97, 99, 101, and 103, according to their number in the valence band complex of the FRFP calculation counted from below. Following the line $\Gamma$–-M in Fig. \ref{fig2} the crossing points of all bands with the Fermi level are clearly visible. The influence of the relativistic effects is most pronounced along the M--K--$\Gamma$ line, where the upper two bands 101 and 103 remain degenerate in the SR calculation along the L-H-A line so that we expect sizable differences in the extremal Fermi surface cross sections to be discussed in the next section. Also the total DOS shows a measurable difference between the FRPL calclations (see Fig.  \ref{fig3} and the SR ones: $N(E_F)=2.45$ (states/eV$\cdot$ f.u.) vs.\ 2.07 states/eV$\cdot$ f.u. reported in Ref.\ 2. In Figs. \ref{fig4} we have highlighted the orbital character of the relativistic band scheme through the colors and their weight, through the thickness of the bands. The flat bands crossings $E_F$ along the in-plane $\Gamma$--M--K and A--H--L directions are admixtures and dominated almost equally by Pt 5$d$ and As 4$p$ states. Whereas The FRFP calculations for  the Pt $5d$ DOS yield 0.77 states/eV$\cdot$ f.u. which is relatively close to 0.7 states/eV$\cdot$ f.u. reported in Ref.\ 2, However, the As $4p$ contributions differ by a factor of 3: 0.99  states/eV$\cdot$ f.u. vs.\ 0.3 in Ref.\ 2. Physically, this means that in our calculations the covalent Pt-As bonding is more pronounced and in general the Pt-5$d$ As-4$p$ complex is more dominant in our case: 72\% vs.\ 48\% . In addition, the bands with an admixture of As 4$p$ and Pt 5$d$ states split along the $\Gamma$--M and the A--L directions.

\section{Fermi Surface Sheets}
\label{sec:4}
\begin{figure*} 
\includegraphics[width=.75\textwidth]{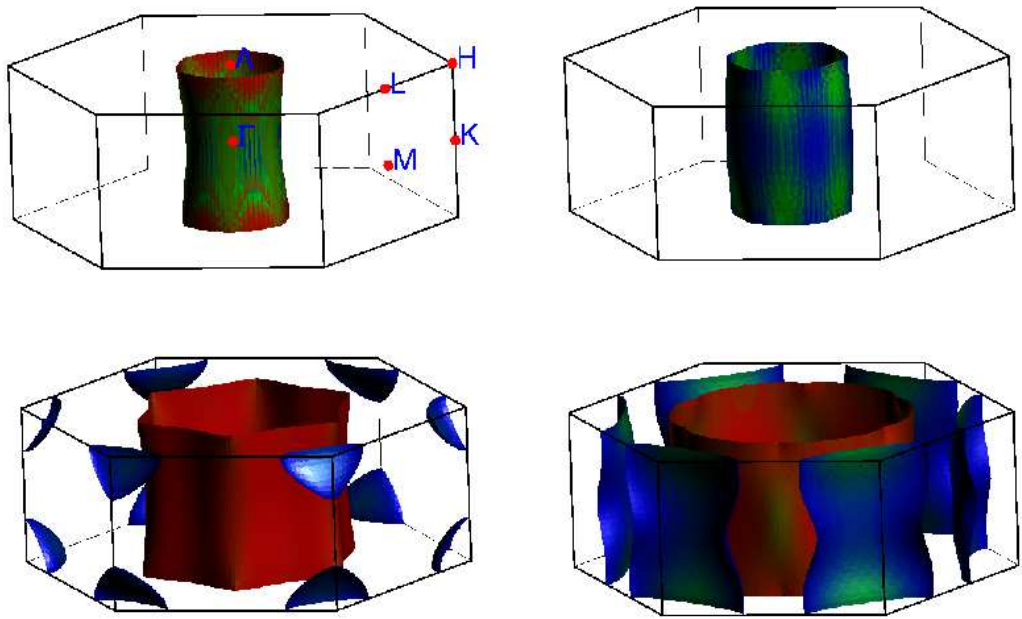}
\caption{ (Color online) Calculated Fermi surface of SrPtAs with extremal orbits indicated. Red, green, and blue color corresponds to fasti, intermediate, and slow electrons, respectively.}
\label{fig5}

\end{figure*}

A detailed insight into the electronic structure can be gained from de Haas--van Alphen (dHvA) measurements which have not  been reported on SrPtAs to date. To initiate possible experiments, we may provide interested readers with these dHvA frequencies and their angular dependencies. The extremal orbits have been calculated using the numerical scheme presented in \cite{Oppeneer}, which is outlined in detail in earlier work on this subject \cite{Yamada}.  The highly anisotropic Fermi surface of SrPtAs calculated within the relativistic scheme (as displayed in Fig. \ref{fig5} consists of six sheets existing in four twofold Kramers degenerated bands (97, 99, 101, 103) and eleven extremal orbits for a magnetic field in c-direction. 
The FS sheets consist of four small tubes directed along $\Gamma$-–A direction, deformed tube-like along the H--K direction, and a small pocket around K. The electrons on the tubes of the third and fourth bands (sheets 5--6 and 10--11) have the largest Fermi velocity while electrons on the first and second bands have a slow velocity but the minimal velocity appears within the sheets around K and H. 

The FS averaged Fermi velocity within the honeycomb-plane amounts to $v_{F,ab}=3.41\times10^7$cm/s in FRFP and $v_{F,ab}=3.54\times10^7$cm/s in SR calculations and the unscreened in-plane plasma frequency  $\Omega^{LDA}_{pl}= 5.49$ eV in SR and $\Omega^{LDA}_{pl}= 5.24$eV in FRFP calculations. For the $c$-axis we found   $\Omega^{LDA}_{pl}=1.105$ eV, only, which gives a direct measure for the strong electronic anisotropy. Within an effective one band Ginzburg--Landau model a huge mass anisotropy of about 22.5 could be estimated. Taking into account an expected mass renormalization due to the high-energy electron--electron interaction by a factor of about 3 as for typical transition metals, one expects $\Omega_{pl}=3.25$ eV. Then the anisotropy could be reduced to about 7.5. The renormalized in-plane plasma frequency  determines the slope of the high-temperature resistivity $\propto \lambda /\Omega^2_{pl}$ and the penetration depth or the condensate density. The calculated Fermi velocities  are relatively large. They slightly exceed those for the sister compound SrPt$_2$As$_2$ \cite{Shein11a}.Using these Fermi velocities, the experimental data of Ref.\ 1 for the slope of the upper critical field near $T_c$ of about 0.096 T/K one estimates within the one-band WHH-theory $H_{c2}(0)\approx 0.158$T, which is at variance with the linear estimation adopted in Ref.\cite{Kudo11} yielding 0.22 T. From Eliashberg-theory for a single effective band \cite{Shulga04} one obtaines 0.14~T in the clean limit adopting  a weak coupling regime for the superconductivity, say, $\lambda=0.6$ in accord with the low $T_c$-value. If, however, $H_{c2}(0)\approx 0.2$ T will be confirmed by future  measurements below 1.8~K , a more sophisticated multiband model should be employed. In this case, $H_{c2}$ should be either dominated by a small group of slow electrons, e.g.\ the pockets around H whereas the slope near $T_c$ is dominated by fast electrons from the large red cylinders (see Fig. \ref{fig5}). 

\section{Conclusions}
 \label{sec:5} 
Here, we have investigated electronic and bonding properties of the superconducting compound SrPtAs. Relativistic effects have been found to change the electronic structure considerably. In particular, the splitting of bands leads to more Fermi surface sheets and extremal orbits to be observed in future de Haas-van Alphen measurements. Our theoretical investigation has shown a band-structure and Fermi surface properties reminiscent of MgB$_2$. However, the covalent bonding between the Pt 5$d$ and As 4$p$ states is much weaker than that between the B states in MgB$_2$ Therefore, the superconductivity occurs at low $T$, only. Also a preliminary analysis of the upper critical field data points to a weak coupling assignment of SrPtAS. In a recent computational investigation by means of LAPW implemented in the WIEN2K code with GGA \cite{Blaha,Perdew2} approximation, the authors claimed that the relativistic effects result mainly in an energy shift and splitting in core and semi-core Pt states which are situated deep under the Fermi level \cite{Shein11}. In contrast, we found a more than three times larger partial density of As derived $4p$-states at the Fermi level within our FRFP scheme and obtained also more bands and FS sheets due to band splitting. Finally, we encourage low temperature measurements such as  specific heat and magnetic quantum oscillations on SrPtAs to further elucidate its electronic properties. Investigations for  SrPtAs on single crystals at ambient and under pressure and as function of doping might provide additional interesting information about the possible role of many-body effects and a deeper insight into the  superconducting pairing mechanism. We foresee that chemical and structural modifications in SrPtAs and related systems may yield optimization of its superconducting properties.
\\
\\
\\

\begin{acknowledgements}
We thank Peter M. Oppeneer, J.\ Mydosh, H.\ Rosner, K.\ Koepernik, and Jeroen van den Brink for discussions. S. Elgazzar thanks the Faculty of Science at the University of Johannesburg for funding of a Postdoctoral Fellowship. AM Strydom acknowledges financial assistance from the SA-NRF (grant 2072956) and DPG (OE511/1-1), and  the DFG (SPP 1458 (D)) is gratefully acknowledged.

{\it Note added in proof.}\\
The results of the present paper have been presented at the
Conference "E-MRS 2011 FALL MEETING, Warsaw (Poland)
in September 19 - 23, 2011. After submission of the manuscript to
the Conference Proceedings we learned about a similar preprint
by S.J. Youn {\it et al.}, arXiv:1202.1604v1 using the different
FLAPW-method. In particular, the calculated
band structure, the plasma frequencies and Fermi velocities are close
each to the other:
5.57 eV and 3.76 $\times10^5$ m/s to be compared with 5.24 eV and 3.54$\times10^5$ m/s in our
calculations. We also obtained a sizable enhancement of the mass
anisotropy by the spin-orbit coupling of 30.8 to be compared to 22.5 in
our case. Thus strong anisotropies of the upper critical field of about
5.55 and  4.74, respectively, are predicted.
\end{acknowledgements}


\begin{thebibliography}{99}
\bibitem{Kudo11}
K.\ Kudo, Y.\ Nishikubo, and M.\ Nohara: J.\ Phys.\ Soc.\ Jpn.\ {\bf 79}, 123710 (2010).
\bibitem{Shein11}
I.R.\ Shein, A.L.\ Ivanovskii, JETP Lett. {\bf 92}, 571 (2010); Physica C, {\bf 471}, 594 (2011).
\bibitem{Shein11a}I.R.\ Shein andA.L.\ Ivanovskii, Phys.\ Rev.\ B 83 (2011) 104501.11
\bibitem{Kakiya11}S.\ Kakiya, K.\ Kudo, Y.\ Nishikubo {\it et al.}, J.\ Phys.\ Soc.\ Jpn.\ {\bf 80}, 093704 (2010).
\bibitem{Ni11}N.\ Ni, J.M.\ Alfred, B.C.\ Chan, and R.J.\ Cava, arXiv:1106.2111v1 (2011).
\bibitem{Cho11}K.\ Cho, M.A.\ Tanatar, H.\ Kim {\it et al.} arXiv:1111.1003v1 (2011).
\bibitem{Neupane11}M.\ Neupane, C.\ Liu, S.-Y.\ Xu {\it et al.}, arXiv:1110.4687v1 (2011).
\bibitem{Shulga04} S.V.\ Shulga and S.-L.\ Drechsler, J.\ Low Temp.\ Phys.\ {\bf 129}, 93 (2002).
\bibitem{Nagamatsu}
J.\ Nagamatsu, N.\ Nakagawa, T.\ Muranaka {\it et al.}, Nature {\bf 410}, 63 (2001).
\bibitem{Opahle}
I.\ Opahle, PhD thesis, University of Technology Dresden, 2001.
\bibitem{Eschrig}
H.\ Eschrig, M.\ Richter, and I.\ Opahle, in Relativistic Electronic Structure Theory Part II: Applications, edited by P.\ Schwerdtfeger (Elsevier, Amsterdam, 2004), pp. 723-776.
\bibitem{Koepernik}
K.\ Koepernik and H.\ Eschrig, Phys.\ Rev.\ B {\bf 59},  1743 (1999).
\bibitem{Perdew1}
J.P.\ Perdew, and Y.\ Wang, Phys.\ Rev.\ B {\bf 45}, 13244 (1992).
 \bibitem{Yoshihiro}
 Y.\ Nishikubo, K.\  Kudo, and M.\  Nohara; J.\ Phys.\ Soc.\ Jpn,\ {\bf 80},  055002 (2011).
\bibitem{Oppeneer}
P.M.\ Oppeneer  and A.J.\ Lodder, J. Phys. F: Met. Phys. {\bf 17}, 1901 (1987).
\bibitem{Yamada}
H.\ Yamada, Physica B {\bf 149}, 390 (1988).
\bibitem{Blaha}
P.\ Blaha, K. Schwarz, G.K.H. Madsen {\it et al.} WIEN2k, An Augmented Plane Wave Plus Local Orbitals Program for Calculating Crystal Properties, Vienna University of Technology, Vienna, 2001.
\bibitem{Perdew2}
J.P.\ Perdew, S.\ Burke, M.\ Ernzerhof, Phys.\ Rev.\ Lett.\ 77 (1996) 3865.
\end{thebibliography}
\end{document}